# Enhanced molecular orientation via NIR-delay-THz scheme: Experimental results at room temperature

Ran Damari[1,2], Amit Beer[1,2], Eli Flaxer[1,3] and Sharly Fleischer[1,2*]

[1]Raymond and Beverly Sackler Faculty of Exact Sciences, School of Chemistry, Tel Aviv University, Tel Aviv 6997801, Israel
[2]Tel-Aviv University center for Light-Matter-Interaction, Tel Aviv 6997801, Israel
[3]AFEKA – Tel-Aviv Academic College of Engineering, 69107 Tel-Aviv, Israel
*email address: sharlyf@tauex.tau.ac.il

**Abstract:** THz fields induce orientation in gas phase molecules via resonant dipole-field interaction. The degree of orientation however remains severely limited due to the practical shortage of high THz-field amplitudes. In this paper, we experimentally demonstrate a concerted Near-IR and THz excitation scheme that provides significant increase in the degree of orientation at room temperature gas ensembles. The experimental results are supported by theoretical simulations and a detailed discussion of the multiple coherent transition pathways involved in the scheme is presented.

## Introduction:

The last four decades have witnessed vast advancements in laser-controlled angular distributions and dynamics of gas phase molecules. A pivotal effort in recent years is aimed at orienting polar gas molecules, such that their permanent dipoles are preferably directed toward a specific lab-frame vector direction (either 'up' or 'down'). Unlike the well-established molecular 'alignment', that relies on the interaction of intense laser pulses with the polarizability tensor of molecules [1–5], 'orientation' is induced via dipole-interaction with dc [6–10] or resonant terahertz (THz) fields[11–16], or via mixed field excitation ($\omega + 2\omega$) through the molecular hyperpolarizability[17–22]. A key feature of transient orientation is the lifted inversion symmetry of the oriented medium. That enables nonlinear optical signatures of even orders in the electric field such as second harmonic generation (SHG) [23–25] and higher[18,26].

In this work we present experimental results of enhanced molecular orientation at room temperature via concerted rotational excitation by Near-IR (NIR) and THz fields with judicious delay apart. All–optical detection of the orientation responses relies on the MOISH technique[24,25], by detecting the SHG of a probe pulse that interacts with the oriented gas.

We start with a short introduction to THz-induced orientation (section I), describe the NIR-delay-THz approach followed by experimental results and theoretical analysis (section II) and experimentally demonstrate enhanced orientation using the developed method (section III).

## Section I: THz-induced molecular orientation

Single-cycle THz fields have been shown to induce molecular orientation in polar gas molecules via resonant field-dipole excitation[13,15,27,28]. With the available THz peak amplitudes ~1MV/cm on a laboratory table-top, the degree of orientation that may be achieved is on the order of few percent, i.e. $\langle\langle cos\theta \rangle\rangle \sim 0.05$ where $\theta$ is the angle between the molecular dipole vector and the THz field polarization direction. The double-brackets stands for ensemble averaging. What compromises the extent to which a thermal ensemble can be controlled (toward orientation in this case) is its finite temperature, i.e. one would like the interaction energy to significantly exceed the inherent thermal energy of the system. This may be achieved by cooling the ensemble prior to its interaction with the laser (in some cases down to the single populated rotational level) and has shown to dramatically increase the degree of alignment and orientation[8,10,29] with the help of mixed dc+NIR fields. In THz-induced orientation however, rotational cooling does not suffice as shown in Fig.1, where we simulated the THz-induced orientation in carbonyl-sulfide (OCS) for varying initial temperatures with fixed THz-field parameters. The THz field is taken as a Gaussian temporal envelope with full width half maximum (FWHM) of 1.2ps and carrier frequency $\omega_0 = 0.5\,THz$, corresponding to our typical experimental field. The THz field amplitudes considered throughout this paper are set to induce orientation of $\langle\langle cos\theta\rangle\rangle \leq 0.01$ in methyl-iodide gas at room temperature ($CH_3I$ with rotational constants: $B = 0.25\,cm^{-1}, D = 2.1 \cdot 10^{-7}\,cm^{-1}$). For simulation details see supplemental information (SI.1).

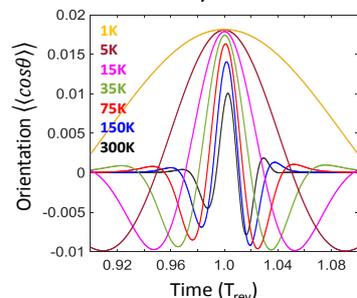

Figure 1: Simulated THz-induced orientation of $CH_3I$ at varying initial temperatures following interaction with a single-cycle THz pulse ($\omega_0 = 0.5\,THz$, $FWHM = 1.2ps$, $E_{THz}$ amplitude $\sim 150kV/cm$).

Fig.1 depicts the calculated THz-induced degree of orientation of methyl-iodide gas ($CH_3I$) for varying initial temperatures. As can be seen, the degree of orientation is hardly affected by the sample temperature, with $\langle\langle cos\theta \rangle\rangle \sim 0.01 \rightarrow 0.017$ for the 300K$\rightarrow$1K range. This is because the resonant THz-field interaction requires spectral overlap between the transition frequency spectrum of the molecules and the THz pulse spectrum. As the temperature decreases, the system becomes more controllable, however in parallel, the populated rotational states and corresponding transition frequencies of the molecular rotors gradually shift towards lower THz frequencies and the spectral overlap between the two interacting entities is gradually decreasing. These two effects seem to balance each other and the degree of orientation remains low. Note the slight change in the temporal line-shape at different temperatures, a result of the relatively large centrifugal distortion coefficient of $CH_3I$ ($D = 2.1 \cdot 10^{-7}\,cm^{-1}$).

In order to tackle this difficulty, Egodapitiya et.al.[30] have demonstrated experimentally that pre-excitation by a near-IR pulse of rotationally cold OCS molecules (~2K) can prepare the ensemble for its efficient interaction with the THz field to further enhance the degree of orientation[31,32]. Ren et.al demonstrated a similar approach with a two-color pulse replacing for the THz field[33]. The NIR-delay-THz excitation scheme was recently revisited theoretically by Tutunnikov et.al [34], where enhanced orientation was predicted feasible at high temperatures as well. This simplifies the practical utilization of the NIR-delay-THz scheme, extending its implementation to a broad range of initial temperatures and gas densities. In this work we experimentally validate the NIR-delay-THz scheme at ambient temperature (300K) using the all-optical MOISH detection [24,25].

**Section II: NIR-delay-THz scheme**

The experimental scheme used in this work includes three pulses with controlled delay apart. The first pulse that interacts with the sample is a 100fs NIR ($\lambda = 800nm$) pulse that induces coherent rotational dynamics via interaction with the anisotropic polarizability tensor of the molecules. The next pulse that interacts with the (already excited) ensemble is a single-cycle THz pulse that induces orientation via resonant dipole-field interaction. The third pulse is a NIR probe that interacts with the oriented ensemble and give rise to SH signal ($\lambda_{signal} = 400nm$) detected by a photomultiplier (PMT) and recorded on a lab-PC. The experimental setup used in this work was reported in our recent paper[25], and is provided in the Supplemental Information section SI.2 for the convenience of the readers. We stress the difference in time-ordering of the pulses: while the THz-delay-NIR scheme of [25] gives rise to a rephasing orientation echo signal, the NIR-delay-THz scheme presented here, results in enhanced non-rephasing molecular orientation. An important aspect of our detection scheme is its relatively short length of interaction, primarily dictated by the Rayleigh range of the NIR pulse. The latter relieves the phase-matching constraints for SHG and enables measurements of a broad range of gas densities. In addition, the orientation signal of interest is temporally separated from orientation responses induced by the THz pulse solely (orientation revivals). This reassures that the SH signal is radiated at THz-field-free conditions, avoiding possible contributions from instantaneous electronic responses induced by the THz field[24].

## *Experimental results:*

Figure 2a depicts the experimental signals induced by the NIR-delay-THz excitation scheme. The time axis is given in units of the rotational revival period of $CH_3I$ ($T_{rev} = 1/2B = 66.67ps$)[25,35]. An ultrashort (100fs) NIR pulse is applied at t=-0.22$T_{rev}$ followed by a single-cycle THz field at t=0. A NIR probe pulse is scanned through the entire timedomain of Fig.2 and its SH signal (400nm) is recorded. The first SH signal is observed at t=0, where the probe pulse temporally overlaps the incident THz field. This signal results from two contributions to the nonlinear susceptibility $\chi^{(2)}$: an instantaneous orientation of the gas and an instantaneous THz-induced electronic response. The signal observed at t=1$T_{rev}$ is quantum revival primarily contributed by the orientation of the gas (marked as Orientation Revival). Those signals are solely induced by the THz pulse as recently reported in [24]. The signal of interest in this work (SOI) is observed at t=0.58$T_{rev}$ and is induced by the action of BOTH the NIR (red) and THz (blue) pulses. This signal precedes the quantum rotational revival by 2Δt (i.e. at $t = 1T_{rev} - 2\Delta t$) and indicates on the transient orientation of the gas. While the t=0 and t=1$T_{rev}$ signals are somewhat 'smeared' as evident by the compromised depth of modulation of these signals, the shape of the SOI is in perfect agreement with the simulated results. The reason for that stems from the significantly short mutual interaction length of the NIR and THz pulses (governed by the Rayleigh length of the NIR beam), as compared to the long interaction length of the t=0 and t=1$T_{rev}$ signals (governed by the THz beam) making them amenable to the ramifications of phase-mismatch and averaging over the Gouy phase of the THz field[25]. Figures 2b,c depict the dependence of the SOI on the excitation pulses. We plot the peak intensity of the SOI as a function of the NIR intensity ($I_{NIR}$ in Fig.2b) and the THz amplitude ($E_{THz}$ in Fig.2c) and find quadratic dependence on both $\left(Signal \propto I_{NIR}^2, E_{THz}^2\right)$. Accounting for the heterodyne nature of the MOISH signal, $Signal \propto (\chi^{(2)})^2 \propto (\langle\langle cos\theta\rangle\rangle)^2$, we deduce that the orientation SOI increases linearly both with the THz field and NIR intensity. This guides our following discussion on the transition pathways that govern the orientation SOI.

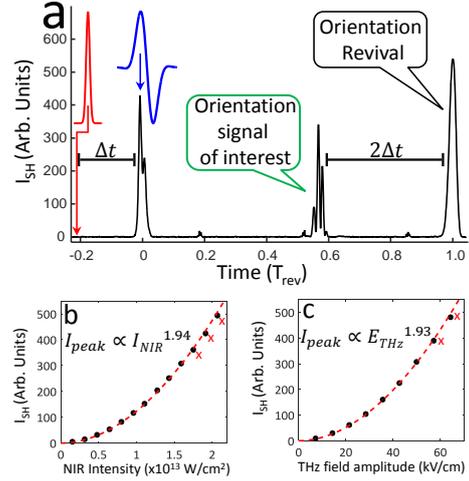

Figure 2: (a) representative experimental signal from $CH_3I$ gas, induced by a NIR pulse and a single-cycle THz field with a delay of $\Delta t = 0.22 T_{rev}$ apart. The orientation signal of interest (SOI) is detected at $0.56 T_{rev}$, i.e. precedes the orientation revival by $2\Delta t$. (b and c) dependence of the SOI signal on the NIR intensity and the THz field amplitude respectively. The quadratic dependence of $I_{SH}$ on both the $I_{NIR}$ and $E_{THz}$ in the homodyne detection indicates that the SOI is induced by one Raman interaction with the NIR and one dipole interaction with the THz. (Data points marked by a red x were excluded from the power fit. When included, the power argument reduces to ~1.87.)

Our first task is set to explain the time of appearance of the SOI signal at a delay of $1T_{rev} - 2\Delta t$ following the THz field (where $t_{NIR} = -\Delta t$ and $t_{THz} = 0$ as depicted in Fig.2a). We start by considering the orientation dynamics induced by the THz field alone. Figure 3a is a graphical presentation of the light-induced transition pathways within the coherent rotational manifold. This presentation approach, inspired by the density matrix formalism, has proven very instrumental in [25,36,37] for tracking multiple interfering pathways within the multi-level rotational system and is used here for the orientation SOI.

### *Orientation by a single THz field:*

We start with a thermal ensemble represented by the grey dots along the main diagonal of the density matrix in Fig.3a. Those are thermally populated rotational states $\rho_{|J\rangle\langle J|} = Z^{-1}\exp[-\frac{E_J}{kT}]$ where $E_J = hBc \cdot J \cdot (J+1)$ are the eigenenergies, $J$ is the rotational quantum number, $h$ is Planck's constant, $B$ is the rotational coefficient in $[cm^{-1}]$, $c$ the speed of light and $Z = \sum_0^\infty (2J+1) \cdot \exp[-\frac{E_J}{kT}]$ is the partition function. A THz field interacts with the thermal ensemble $V_{THz} = -\vec{\mu} \cdot \vec{E}_{THz} = -\mu E_{THz}\cos\theta$ and induces one quantum coherences (1QC) that are situated along the first off-diagonal of the density matrix, $|J\rangle\langle J+1|$ (striped circles). The green arrows mark the transition pathways where the solid green arrows induce $|J\rangle\langle J| \to |J\rangle\langle J+1|$ and dashed green arrows induce $|J+1\rangle\langle J+1| \to |J\rangle\langle J+1|$ operating on the $\langle bra|$ and $|ket\rangle$ respectively. The magnitudes of the 1QC terms clearly depend on $E_{THz}$, on the coupling between levels $\langle J,m|\cos\theta|J+1,m\rangle$ and on the difference between adjacent population terms $\rho_{|J\rangle\langle J|} - \rho_{|J+1\rangle\langle J+1|}$ that is dictated by the initial temperature of the ensemble. In fact, the solid and dashed transition vectors (green) interfere destructively, hence the net 1QC amplitudes are severely reduced as the temperature increases. Once created by the THz field, the 1QCs accumulate phase throughout their field-free evolution and manifest in periodic molecular orientation events with every T$_{rev}$. The black curve in Fig.3b shows a simulation of the THz-induced orientation of the gas throughout the first revival period. The dashed red and solid blue curves show the orientation calculated for the $|J_{even}\rangle\langle J_{even}+1|$ (1QC$_{even}$) and $|J_{odd}\rangle\langle J_{odd}+1|$ (1QC$_{odd}$) selectively, demonstrating the inherent destructive interference at t=T$_{rev}$/2 and constructive interference at t=0 and at t=T$_{rev}$.

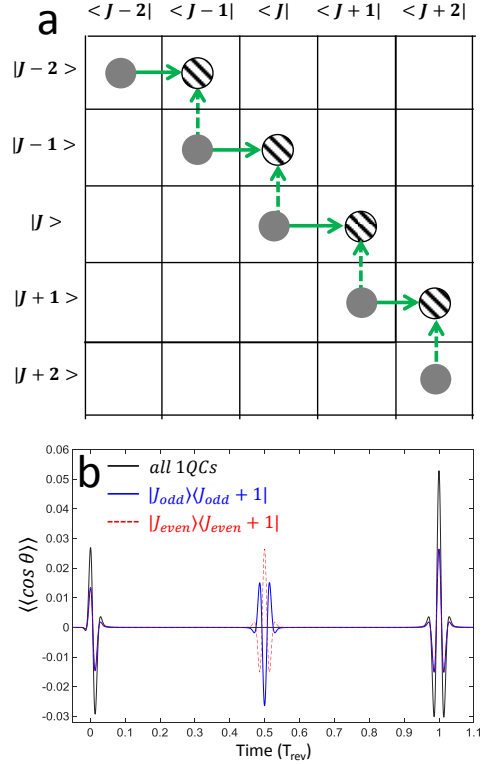

Figure 3: (a) Density matrix representation of the transitions induced by the THz field (see text). (b) Coherent orientation responses of odd (blue) and even (red) coherences selectively and their sum (black).

### *THz-induced orientation in rotationally excited molecules:*

Next, we proceed to analyze the excitation scheme of the SOI in this work. Here, the first pulse that excites the thermal ensemble is a short duration NIR pulse that interacts via $V_{NIR} = -\frac{1}{4}\Delta\alpha|E(t)|^2\cos^2\theta$. The latter induces two quantum rotational coherence (2QC) terms marked by the open circles on the second off-diagonal in Fig.4a.

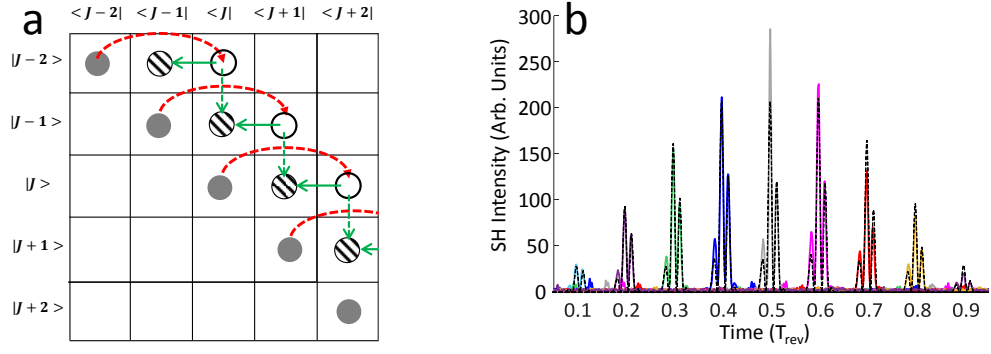

Figure 4: (a) density matrix representation of the NIR-delay-THz transitions that induce the SOI. (b) Experimental (color-coded) and simulated (dashed black) curves of the SOI for fixed $I_{NIR}$ and $E_{THz}$ with different delays apart. The time of incidence of the THz field is set as t=0 and the signals obey $t_{SOI} = 1T_{rev} - 2\Delta t$.

Note that we mark the transition arrows (dashed red) from the population term $|J\rangle\langle J|$ to and $|J\rangle\langle J + 2|$ and discard the arrows that couple $|J + 2\rangle\langle J + 2|$ to $|J\rangle\langle J + 2|$ for the sake of graphical compactness (these destructively interfering pathways are fully accounted for in our simulations). The 2QCs manifest as periodic recurrences of molecular alignment every $T_{rev}/2$, while retaining zero net orientation. At delay $\Delta t$ after the NIR excitation, a single-cycle THz field interacts with the population terms resulting in orientation revivals (as described in the previous section). In addition, the THz field also interacts with the existing 2QCs that were induced by the NIR and projects them (green arrows) to the 1QCs manifold (striped circles). Let us calculate the phases of the 1QCs throughout their journey:

Under field-free evolution, the 2QCs (that were 'born' at $t = -\Delta t$) evolve according to $e^{-i\varphi} = e^{-iH\Delta t/\hbar}$, namely they accumulate phase given by:
$$\varphi_{2QC} = (E_{J+2} - E_J) \cdot \pi \cdot \Delta t = (4J + 6)\pi\Delta t$$
where $\Delta t$ is in units of $T_{rev}(= 1/2B)$. Thus, two adjacent 2QC terms ($|J - 1\rangle\langle J + 1|$ and $|J\rangle\langle J + 2|$) accumulate a phase difference of $\Delta\varphi_{2QC} = 4\pi\Delta t$. If left to evolve unperturbedly, the phase difference between all adjacent 2QCs will be an even multiple of $\pi$ every $T_{rev}/2$. Here however, the 2QCs are projected by the THz field (at t=0, namely at a delay $\Delta t$ after their birth) to the 1QCs manifold where they continue to accumulate phase only at half the rate $\Delta\varphi_{1QC} = 2\pi t_{SOI}$. The total phase accumulated between two adjacent 1QCs is thus $\Delta\varphi_{1QC}(t_{SOI}) = 4\pi\Delta t + 2\pi t_{SOI} = 2\pi(2\Delta t + t_{SOI})$. Correspondingly, the recurrence time for orientation, at which all of the $\Delta\varphi_{1QC}$ are even number of $\pi$, is reached when $(2\Delta t + t_{SOI}) = 1$, namely at $t_{SOI} = 1 - 2\Delta t$ (as found experimentally and theoretically in Fig.2).

Next, we set to explore the dependence of the SOI on $\Delta t$. Figure 4b shows the experimental SOI obtained with different delays $\Delta t$ (color-coded) and fixed $E_{THz}$ and $I_{NIR}$ values. The maximal SOI signal is obtained for $\Delta t = T_{rev}/4$ and the overall trend is agreement with the simulated results (dashed black) and with ref. [30]. We consider the phase difference between the two pathways that interfere to create the 1QC term responsible for orientation. With $\varphi_{|J-1\rangle\langle J+1|} = (4J + 2)\pi/4$ and $\varphi_{|J+2\rangle\langle J|} = (4J + 6)\pi/4$ the phase difference $\Delta\varphi$ between the two is exactly $\pi$, setting the stage for a purely constructive interference between the dashed and solid green transition arrows that lead to each 1QC term. At all other delays ($0 < \Delta t < \frac{T_{rev}}{4}$ and $\frac{T_{rev}}{4} < \Delta t < \frac{T_{rev}}{2}$) the constructive interference between the two pathways gradually decreases as $\Delta t$ moves away from $\frac{T_{rev}}{4}$. Note that while the experimental and simulated results are in excellent agreement, the signal at $\Delta t = \frac{T_{rev}}{4}$ (grey signal at $0.5T_{rev}$ in Fig.4b) seems to exceed the simulated trend-line. This results from constructive interference between the $\langle \cos\theta \rangle$ and $\langle \cos^3\theta \rangle$ contributions, lifting the transient inversion symmetry of the gas medium at this specific delay $\Delta t$. In fact, the contributions of $\langle \cos^3\theta \rangle$ to the lifted inversion symmetry manifest in all of the scans shown in Fig.4b (small signal peaks at the level of ~10 in the arbitrary units of Fig.4b), however their time of appearance overlaps that of $\langle cos\theta \rangle$ only at $\Delta t = \frac{T_{rev}}{4}$.

The above discussion highlights a fundamental difference between the typical THz-induced orientation (observed at T$_{rev}$) and the NIR-delay-THz scheme (that induces the SOI) with respect to the natures of interfering pathways; while the THz-induced orientation is inherently compromised by the destructive interference of transition pathways sourced in adjacent population terms (temperature effect), the NIR-delay-THz scheme enjoys the constructive interferences of adjacent 2QC terms. A uniquely desirable possibility of the latter is the potential enhancement of the SOI degree of orientation, far beyond that induced by the THz-field alone, as demonstrated hereafter.

### *Section IV: Enhanced degree of orientation via the NIR-delay-THz scheme*

The main limiting factor in THz-induced molecular orientation experiments is the practical shortage in THz-field amplitude. In Fig.2b and associated text we have shown that the degree of orientation of the SOI increases linearly with the NIR intensity (I$_{NIR}$). In what follows, we set to demonstrate and characterize the extent to which the orientation can be further enhanced using the NIR-delay-THz scheme. Naturally, we wish to compare between the SOI and the orientation signal induced solely by the THz field (at t=T$_{rev}$). A direct comparison between the two does not suffice, since these signals are radiated from interaction regions of different lengths (with the THz-induced length of interaction larger than that of the SOI). Therefore we seek for a different reference signal that serves as a direct indicator for the THz-induced orientation value. The latter is provided by the recently reported THz-NIR echo signal[25].

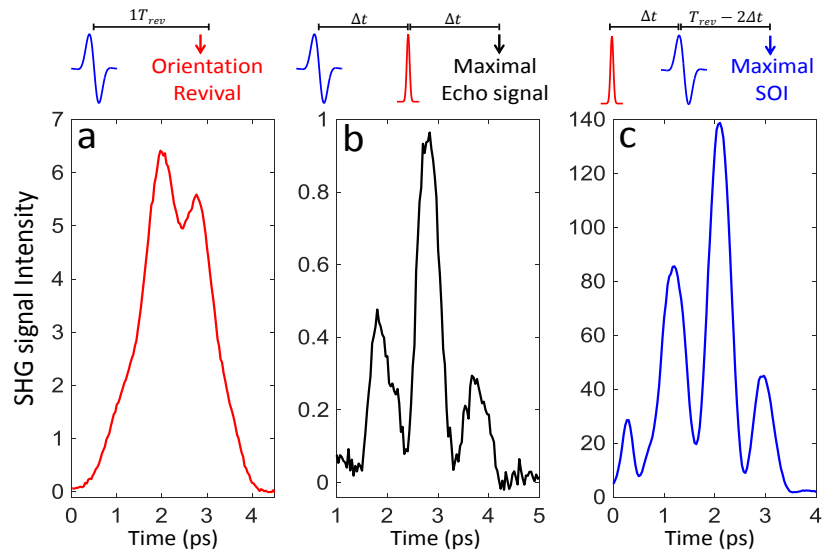

Figure 5: (a) Experimental orientation signal induced by the THz pulse ($E_{THz} \sim 50 kV/cm$) detected at 1T$_{rev}$. (b) Maximal echo signal induced by THz-delay-NIR pulse for orientation calibration purposes (see text). (c) Maximal SOI obtained in this work with the same THz pulse as in (a) and with $I_{NIR} \sim 2 \cdot 10^{13} \ W/cm^2$. Corresponding excitation schemes are shown above each panel respectively.

The THz-NIR echo signal shares the exact same orders of interaction with the THz and NIR pulses (linear with both, similar to the SOI) and is also detected as a SH signal. Thus the THz-NIR echo and our SOI share the exact same effective length of interaction and their SH signals can be directly compared to each other. We utilize the fact that the THz-NIR echo signal is directly related to the THz-induced orientation value as the maximal degree of orientation provided by the echo response is 1/2 that of the (THz-induced) orientation revival [25]. Figure 5 depicts the experimental SH signals of the THz-induced orientation (red, panel a), the maximal echo signal (panel b, black) and the maximal SOI (blue) obtained with a fixed THz-field amplitude and probe intensity.

The red curve in Fig.5a shows the experimental SH signal at the revival of orientation induced solely by the THz field (see excitation scheme above panel a). The peak intensity of the signal is 6.4 in the arbitrary units of the experiment that are shared by all three figure panels. The black curve in Fig.5b depicts the maximal echo signal induce by the same THz field followed by a NIR pulse. This was obtained from a set of THz-delay-NIR measurements performed with varying NIR intensities. The peak SH signal of the orientation echo is found to be 0.96 (in the same arbitrary unit scale). This value corresponds to $1/2$ the degree of orientation of that induced by the THz field alone based on our simulations [25] and therefore expected to be $1/4$ in SH signal. Thus, the obtained ratio of the two SH signals ($0.96/6.4 = 0.15$ instead of $0.25$) results from the different effective interaction lengths of the two schemes as described above. In what follows we consider the value of $0.96$ as the reference for the degree of orientation induced by the THz field alone. Namely, had the THz-induced orientation and the THz-NIR echo signals emanated from the exact same interaction length, the peak orientation signal in Fig.5a would have been $0.96 \times 4 = 3.84$. The blue curve in Fig.5c shows the maximal SOI obtained in our experiment using the NIR-delay-THz scheme with peak SH intensity of $139$. This signal is $\sim 20$ fold larger than that of Fig.5a and $\sim 36$ fold larger than the THz-induced orientation level (for the same length of interaction), indicating an enhancement factor of $\sqrt{36} = 6$ in the degree of molecular orientation $\langle\langle cos\theta \rangle\rangle$. In fact, our simulations predict that significantly larger orientation enhancements are feasible at room temperatures (>18-fold, see Supplemental information section SI.3), if one overcomes the detrimental effects of strong-field ionization of the gas that have set the experimental limit in this work. This can be achieved, for example, by optimizing the NIR pulse duration for maximal 2QCs amplitude (alignment) or with a train of NIR pulses[38] separated by the revival period prior to the THz excitation.

To summarize, we have studied the rotational responses induced in an ensemble of molecular rotors by concerted NIR and THz rotational excitation. The NIR-delay-THz scheme enables significant increase in orientation, far beyond that of the THz field alone. While we have experimentally demonstrated 6-fold enhancement of the orientation at ambient temperature, our theoretical simulations predict significantly larger enhancements if one is able to avoid detrimental strong-field effects of the NIR pulse. The presented scheme is uniquely desirable for the preparation of highly-oriented molecular ensembles that are currently severely limited by the practical availability of high THz field amplitudes.

**Acknowledgements:** The authors acknowledge the support of the Israel Science Foundation (926/18, 1856/22), the Wolfson family foundation (PR/ec/20419) and the PAZI foundation.

**References**

(1)  Stapelfeldt, H.; Seideman, T. Colloquium : Aligning Molecules with Strong Laser Pulses. *Rev. Mod. Phys.* **2003**, *75* (2), 543–557. https://doi.org/10.1103/RevModPhys.75.543.

(2)  Lemeshko, M.; Krems, R. V.; Doyle, J. M.; Kais, S. Manipulation of Molecules with Electromagnetic Fields. *Molecular Physics*. Taylor & Francis Group July 14, 2013, pp 1648–1682. https://doi.org/10.1080/00268976.2013.813595.

(3)  Ohshima, Y.; Hasegawa, H. Coherent Rotational Excitation by Intense Nonresonant Laser Fields. *Int. Rev. Phys. Chem.* **2010**, *29* (4), 619–663. https://doi.org/10.1080/0144235X.2010.511769.

(4)  Fleischer, S.; Khodorkovsky, Y.; Gershnabel, E.; Prior, Y.; Averbukh, I. S. Molecular Alignment Induced by Ultrashort Laser Pulses and Its Impact on Molecular Motion. *Isr. J. Chem.* **2012**, *52* (5). https://doi.org/10.1002/ijch.201100161.

(5) Koch, C. P.; Lemeshko, M.; Sugny, D. Quantum Control of Molecular Rotation. *Rev. Mod. Phys.* **2019**, *91* (3), 035005. https://doi.org/10.1103/RevModPhys.91.035005.

(6) Friedrich, B.; Herschbach, D. R. Spatial Orientation of Molecules in Strong Electric Fields and Evidence for Pendular States. *Nature* **1991**, *353* (6343), 412–414. https://doi.org/10.1038/353412a0.

(7) Friedrich, B.; Herschbach, D. Enhanced Orientation of Polar Molecules by Combined Electrostatic and Nonresonant Induced Dipole Forces. *J. Chem. Phys.* **1999**, *111* (14), 6157–6160. https://doi.org/10.1063/1.479917.

(8) Holmegaard, L.; Nielsen, J. H.; Nevo, I.; Stapelfeldt, H.; Filsinger, F.; Küpper, J.; Meijer, G. Laser-Induced Alignment and Orientation of Quantum-State-Selected Large Molecules. *Phys. Rev. Lett.* **2009**, *102* (2), 023001. https://doi.org/10.1103/PhysRevLett.102.023001.

(9) Goban, A.; Minemoto, S.; Sakai, H. Laser-Field-Free Molecular Orientation. *Phys. Rev. Lett.* **2008**, *101* (1). https://doi.org/10.1103/PhysRevLett.101.013001.

(10) Ghafur, O.; Rouzée, A.; Gijsbertsen, A.; Siu, W. K.; Stolte, S.; Vrakking, M. J. J. Impulsive Orientation and Alignment of Quantum-State-Selected NO Molecules. *Nat. Phys.* **2009**, *5* (4), 289–293. https://doi.org/10.1038/nphys1225.

(11) Hebling, J.; Yeh, K.; Hoffmann, M. C.; Bartal, B.; Nelson, K. A. Generation of High-Power Terahertz Pulses by Tilted-Pulse-Front Excitation and Their Application Possibilities. *J. Opt. Soc. Am. B* **2008**, *25* (7), B6. https://doi.org/10.1364/JOSAB.25.0000B6.

(12) Fleischer, S.; Field, R. W.; Nelson, K. A. Commensurate Two-Quantum Coherences Induced by Time-Delayed THz Fields. *Phys. Rev. Lett.* **2012**, *109* (12), 123603. https://doi.org/10.1103/PhysRevLett.109.123603.

(13) Fleischer, S.; Zhou, Y.; Field, R. W.; Nelson, K. A. Molecular Orientation and Alignment by Intense Single-Cycle THz Pulses. *Phys. Rev. Lett.* **2011**, *107* (16), 163603. https://doi.org/10.1103/PhysRevLett.107.163603.

(14) Shalaby, M.; Hauri, C. P. Air Nonlinear Dynamics Initiated by Ultra-Intense Lambda-Cubic Terahertz Pulses. *Appl. Phys. Lett.* **2015**, *106* (18), 181108. https://doi.org/10.1063/1.4919876.

(15) Damari, R.; Kallush, S.; Fleischer, S. Rotational Control of Asymmetric Molecules: Dipole- versus Polarizability-Driven Rotational Dynamics. *Phys. Rev. Lett.* **2016**, *117* (10), 103001. https://doi.org/10.1103/PhysRevLett.117.103001.

(16) Damari, R.; Rosenberg, D.; Fleischer, S. Coherent Radiative Decay of Molecular Rotations: A Comparative Study of Terahertz-Oriented versus Optically Aligned Molecular Ensembles. *Phys. Rev. Lett.* **2017**, *119* (3), 033002. https://doi.org/10.1103/PhysRevLett.119.033002.

(17) De, S.; Znakovskaya, I.; Ray, D.; Anis, F.; Johnson, N. G.; Bocharova, I. A.; Magrakvelidze, M.; Esry, B. D.; Cocke, C. L.; Litvinyuk, I. V; Kling, M. F. Field-Free Orientation of CO Molecules by Femtosecond Two-Color Laser Fields. *Phys. Rev. Lett.* **2009**, *103* (15), 153002. https://doi.org/10.1103/PhysRevLett.103.153002.

(18) Frumker, E.; Hebeisen, C. T.; Kajumba, N.; Bertrand, J. B.; Wörner, H. J.; Spanner, M.; Villeneuve, D. M.; Naumov, A.; Corkum, P. B. Oriented Rotational Wave-Packet Dynamics Studies via High Harmonic Generation. *Phys. Rev. Lett.* **2012**, *109* (11), 113901. https://doi.org/10.1103/PhysRevLett.109.113901.

(19) Lin, K.; Tutunnikov, I.; Qiang, J.; Ma, J.; Song, Q.; Ji, Q.; Zhang, W.; Li, H.; Sun, F.; Gong, X.; Li, H.; Lu, P.; Zeng, H.; Prior, Y.; Averbukh, I. S.; Wu, J. All-Optical Field-Free Three-Dimensional Orientation of Asymmetric-Top Molecules. *Nat. Commun.* **2018**, *9* (1), 1–9. https://doi.org/10.1038/s41467-018-07567-2.

(20) Spanner, M.; Patchkovskii, S.; Frumker, E.; Corkum, P. Mechanisms of Two-Color Laser-Induced Field-Free Molecular Orientation. *Phys. Rev. Lett.* **2012**, *109* (11), 113001. https://doi.org/10.1103/PhysRevLett.109.113001.

(21) Kraus, P. M.; Tolstikhin, O. I.; Baykusheva, D.; Rupenyan, A.; Schneider, J.; Bisgaard, C. Z.; Morishita, T.; Jensen, F.; Madsen, L. B.; Wörner, H. J. Observation of Laser-Induced Electronic Structure in Oriented Polyatomic Molecules. *Nat. Commun.* **2015**, *6* (1), 7039. https://doi.org/10.1038/ncomms8039.

(22) Wu, J.; Zeng, H. Field-Free Molecular Orientation Control by Two Ultrashort Dual-Color Laser Pulses. *Phys. Rev. A* **2010**, *81* (5), 053401. https://doi.org/10.1103/PhysRevA.81.053401.

(23) Beer, A.; Hershkovitz, D.; Fleischer, S. Iris-Assisted Terahertz Field-Induced Second Harmonic Generation in Air. *arXiv* **2019**, *3*. https://doi.org/10.1364/ol.44.005190.

(24) Beer, A.; Damari, R.; Chen, Y.; Fleischer, S. Molecular Orientation-Induced Second-Harmonic Generation: Deciphering Different Contributions Apart. *J. Phys. Chem. A* **2022**, *126* (23), 3732–3738. https://doi.org/10.1021/acs.jpca.2c03237.

(25) Damari, R.; Beer, A.; Fleischer, S. Orientation Echoes via Concerted Terahertz and Near-IR Excitation. **2022**. https://doi.org/10.48550/arxiv.2208.02016.

(26) Kraus, P. M.; Rupenyan, A.; Wörner, H. J. High-Harmonic Spectroscopy of Oriented OCS Molecules: Emission of Even and Odd Harmonics. *Phys. Rev. Lett.* **2012**, *109* (23), 233903. https://doi.org/10.1103/PhysRevLett.109.233903.

(27) Hong, Q.-Q.; Fan, L.-B.; Shu, C.-C.; Henriksen, N. E. Generation of Maximal Three-State Field-Free Molecular Orientation with Terahertz Pulses. *Phys. Rev. A* **2021**, *104* (1), 013108. https://doi.org/10.1103/PhysRevA.104.013108.

(28) Shu, C.-C.; Henriksen, N. E. Field-Free Molecular Orientation Induced by Single-Cycle THz Pulses: The Role of Resonance and Quantum Interference. *Phys. Rev. A* **2013**, *87* (1), 013408. https://doi.org/10.1103/PhysRevA.87.013408.

(29) Nevo, I.; Holmegaard, L.; Nielsen, J. H.; Hansen, J. L.; Stapelfeldt, H.; Filsinger, F.; Meijer, G.; Küpper, J. Laser-Induced 3D Alignment and Orientation of Quantum State-Selected Molecules. *Phys. Chem. Chem. Phys.* **2009**, *11* (42), 9912. https://doi.org/10.1039/b910423b.

(30) Egodapitiya, K. N.; Li, S.; Jones, R. R. Terahertz-Induced Field-Free Orientation of Rotationally Excited Molecules. *Phys.*

*Rev. Lett.* **2014**, *112* (10), 103002. https://doi.org/10.1103/PhysRevLett.112.103002.

(31) Yoshida, M.; Ohtsuki, Y. Control of Molecular Orientation with Combined Near-Single-Cycle THz and Optimally Designed Non-Resonant Laser Pulses: Carrier-Envelope Phase Effects. *Chem. Phys. Lett.* **2015**, *633* (1), 169–174. https://doi.org/10.1016/j.cplett.2015.05.041.

(32) Kitano, K.; Ishii, N.; Itatani, J. High Degree of Molecular Orientation by a Combination of THz and Femtosecond Laser Pulses. *Phys. Rev. A* **2011**, *84* (5), 053408. https://doi.org/10.1103/PhysRevA.84.053408.

(33) Ren, X.; Makhija, V.; Li, H.; Kling, M. F.; Kumarappan, V. Alignment-Assisted Field-Free Orientation of Rotationally Cold CO Molecules. *Phys. Rev. A - At. Mol. Opt. Phys.* **2014**, *90* (1), 013419. https://doi.org/10.1103/PHYSREVA.90.013419/FIGURES/6/MEDIUM.

(34) Tutunnikov, I.; Xu, L.; Prior, Y.; Averbukh, I. S. Echo-Enhanced Molecular Orientation at High Temperatures. **2022**. https://doi.org/10.48550/arxiv.2207.08274.

(35) Rosenberg, D.; Damari, R.; Kallush, S.; Fleischer, S. Rotational Echoes: Rephasing of Centrifugal Distortion in Laser-Induced Molecular Alignment. *J. Phys. Chem. Lett.* **2017**, *8* (20), 5128–5135. https://doi.org/10.1021/acs.jpclett.7b02215.

(36) Rosenberg, D.; Damari, R.; Fleischer, S. Echo Spectroscopy in Multilevel Quantum-Mechanical Rotors. *Phys. Rev. Lett.* **2018**, *121* (23), 234101. https://doi.org/10.1103/PhysRevLett.121.234101.

(37) Wang, P.; He, L.; He, Y.; Sun, S.; Liu, R.; Wang, B.; Lan, P.; Lu, P. Multilevel Quantum Interference in the Formation of High-Order Fractional Molecular Alignment Echoes. *Opt. Express* **2021**, *29* (2), 663. https://doi.org/10.1364/OE.411218.

(38) Cryan, J. P.; Bucksbaum, P. H.; Coffee, R. N. Field-Free Alignment in Repetitively Kicked Nitrogen Gas. *Phys. Rev. A* **2009**, *80* (6), 063412. https://doi.org/10.1103/PhysRevA.80.063412.